\documentclass[12pt,preprint]{aastex}
\usepackage{amsmath}
\usepackage{pdfpages}
\usepackage{graphicx}

\shorttitle{Models of Rosetta/OSIRIS phase function}
\shortauthors{Moreno et al.}

\begin{document}


\title{Models of Rosetta/OSIRIS 67P dust coma phase function}  


\author{F. Moreno 
\affil{Instituto de Astrof\'\i sica de Andaluc\'\i a, CSIC,
 c/ Glorieta de la Astronom\'\i a s/n, 18008 Granada, Spain}
\email{fernando@iaa.es}}

\author{D. Guirado  
\affil{Instituto de Astrof\'\i sica de Andaluc\'\i a, CSIC,
 c/ Glorieta de la Astronom\'\i a s/n, 18008 Granada, Spain}}

\author{O. Mu\~noz   
\affil{Instituto de Astrof\'\i sica de Andaluc\'\i a, CSIC,
  c/ Glorieta de la Astronom\'\i a s/n, 18008 Granada, Spain}}

\author{I. Bertini
\affil{Department of Physics and Astronomy `G. Galilei',
  University of Padova, Vicolo dell' Osservatorio 3, I-35122 Padova,
  Italy}}

\author{C. Tubiana 
\affil{Max-Planck-Institut f\"ur Sonnensystemforschung,
  Justus-von-Liebig-Weg, 3, D-37077 G\"ottingen, Germany}}

\author{C. G\"uttler 
\affil{Max-Planck-Institut f\"ur Sonnensystemforschung,
  Justus-von-Liebig-Weg, 3, D-37077 G\"ottingen, Germany}}

\author{M. Fulle  
\affil{Osservatorio Astronomico, Via Tiepolo 11, I-34143 Trieste, Italy }}

\author{A. Rotundi 
\affil{INAF - Istituto di Astrofisica e Planetologia Spaziali, Via
  Fosso del Cavaliere, 100, I-00133 Rome, Italy} 
\affil{Universit\'a degli Studi di Napoli Parthenope, Dip. di Scienze e
  Tecnologie, CDN IC4, I-80143 Naples, Italy}}

\author{V. Della Corte
\affil{INAF - Istituto di Astrofisica e Planetologia Spaziali, Via
  Fosso del Cavaliere, 100, I-00133 Rome, Italy} 
\affil{Universit\'a degli Studi di Napoli Parthenope, Dip. di Scienze e
  Tecnologie, CDN IC4, I-80143 Naples, Italy}}

\author{S.L. Ivanovski 
\affil{INAF - Istituto di Astrofisica e Planetologia Spaziali, Via
  Fosso del Cavaliere, 100, I-00133 Rome, Italy}}

\author{G. Rinaldi  
\affil{INAF - Istituto di Astrofisica e Planetologia Spaziali, Via
  Fosso del Cavaliere, 100, I-00133 Rome, Italy}}

\author{D. Bockel\'ee-Morvan  
\affil{LESIA, Observatoire de Paris, PSL Research University, CNRS,
  Sorbonne Universit\'es, UPMC Univ. Paris 06, Univ. Paris-Diderot, 
Sorbonne Paris Cit\'e, 5 place Jules Janssen, F-92195 Meudon, France}}

\author{V.V. Zakharov 
\affil{Laboratoire de M\'et\'eorologie Dynamique, 
UPMC, Sorbonne Universit\'es, Paris, France}
\affil{INAF - Istituto di Astrofisica e Planetologia Spaziali, Via
  Fosso del Cavaliere, 100, I-00133 Rome, Italy}}

\author{J. Agarwal 
\affil{Max-Planck-Institut f\"ur Sonnensystemforschung,
  Justus-von-Liebig-Weg, 3, D-37077 G\"ottingen, Germany}}

\author{S. Mottola  
\affil{Deutsches Zentrum f\"ur Luft- und Raumfahrt (DLR), Institut
  f\"ur Planetenforschung, Rutherfordstra{\ss}e 23, 12489 Berlin,
  Germany}}

\author{I. Toth
\affil{MTA CSFK Konkoly Observatory of the Hungarian Academy of Sciences,
  Konkoly Thege M. ut 15-17. H1121 Budapest, Hungary}}

\author{E. Frattin 
\affil{Department of Physics and Astronomy `G. Galilei',
  University of Padova, Vicolo dell' Osservatorio 3, I-35122 Padova,
  Italy}}

\author{L.M. Lara   
\affil{Instituto de Astrof\'\i sica de Andaluc\'\i a, CSIC,
 c/  Glorieta de la Astronom\'\i a s/n, 18008 Granada, Spain}}

\author{P.J. Guti\'errez 
\affil{Instituto de Astrof\'\i sica de Andaluc\'\i a, CSIC,
 c/ Glorieta de la Astronom\'\i a s/n, 18008 Granada, Spain}}

\author{Z.Y. Lin
\affil{National Central University No.300 Jhongda Rd., Jhongli city,
  Taoyuan County, 320 Taiwan}}

\author{L. Kolokolova
\affil{Planetary Data System Group, Department of Astronomy,
  Rm. 1207D, Atlantic Bldg., University of Maryland, College Park, MD,
  20742, USA}}

\author{
H. Sierks
\affil{ Max Planck Institute for Solar System Research,
  Justus-von-Liebig-Weg 3, 37077 G{\"o}ttingen, Germany}}

\author{
G. Naletto
\affil{University of Padova, Department of Physics and Astronomy
  ``Galileo Galilei'', Via Marzolo 8, 35131 Padova, Italy}
\affil{University of Padova, Center of Studies and Activities for
  Space (CISAS) ``G. Colombo'', Via Venezia 15, 35131 Padova, Italy} 
\affil{CNR-IFN UOS Padova LUXOR, Via Trasea 7, 35131 Padova, Italy}}

\author{
P. L. Lamy 
\affil{Laboratoire Atmosph\`eres, Milieux et Observations Spatiales, 
CNRS \& Universit\'e de Versailles Saint-Quentin-en-Yvelines, 11 
Boulevard d’Alembert, 78280 Guyancourt, France}}

\author{
R. Rodrigo
\affil{Centro de Astrobiolog\'\i a, CSIC-INTA, 28850 Torrej\'on de Ardoz,
  Madrid, Spain}
\affil{International Space Science Institute, Hallerstrasse 6, 3012
  Bern, Switzerland}}

\author{
D. Koschny
\affil{Science Support Office, European Space Research and Technology
  Centre/ESA, Keplerlaan 1, Postbus 299, 2201 AZ Noordwijk ZH, The
  Netherland}}

\author{
B. Davidsson
\affil{ Jet Propulsion Laboratory, M/S 183-401, 4800 Oak Grove Drive,
  Pasadena, CA 91109, USA}}

\author{
M. A. Barucci
\affil{LESIA, Observatoire de Paris, PSL Research University, CNRS,
  Univ. Paris Diderot, Sorbonne Paris Cit\'e, UPMC Univ. Paris 06,
  Sorbonne Universit\'es, 5 place Jules Janssen, 92195 Meudon, France}} 

\author{
J.-L. Bertaux 
\affil{Laboratoire Atmosph\`eres, Milieux et Observations Spatiales, 
CNRS \& Universit\'e de Versailles Saint-Quentin-en-Yvelines, 11 
Boulevard d’Alembert, 78280 Guyancourt, France}}

\author{
D. Bodewits 
\affil{Department of Astronomy, University of Maryland, College Park,
  MD 20742-2421, USA}} 

\author{
G. Cremonese 
\affil{INAF, Astronomical Observatory of Padova, Vicolo
  dell'Osservatorio 5, 35122 Padova, Italy}} 

\author{
V. Da Deppo 
\affil{CNR-IFN UOS Padova LUXOR, Via Trasea 7, 35131 Padova, Italy}}

\author{
S. Debei
\affil{University of Padova, Department of Industrial Engineering, Via
  Venezia 1, 35131 Padova, Italy}} 

\author{
M. De Cecco 
\affil{ University of Trento, Faculty of Engineering, Via Mesiano 77,
  38121 Trento, Italy}} 

\author{
J. Deller  
\affil{Max Planck Institute for Solar System Research,
  Justus-von-Liebig-Weg 3, 37077 G{\"o}ttingen, Germany}}

\author{
S. Fornasier 
\affil{LESIA, Observatoire de Paris, PSL Research University, CNRS,
  Univ. Paris Diderot, Sorbonne Paris Cit\'e, UPMC Univ. Paris 06,
  Sorbonne Universit\'es, 5 place Jules Janssen, 92195 Meudon, France}}

\author{
W.-H. Ip
\affil{ Graduate Institute of Astronomy, National Central University,
  300 Chung-Da Rd, Chung-Li 32054, Taiwan}
\affil{Space Science Institute, Macau University of Science and
  Technology, Avenida Wai Long, Taipa, Macau}}

\author{
H. U. Keller
\affil{Institut f\"ur Geophysik und extraterrestrische Physik,
  Technische Universit\"at Braunschweig, Mendelssohnstr. 3, 38106
  Braunschweig, Germany}
\affil{Deutsches Zentrum f\"ur Luft- und Raumfahrt (DLR), Institut
  f\"ur Planetenforschung, Rutherfordstra{\ss}e 23, 12489 Berlin,
  Germany}}

\author{
M. Lazzarin
\affil{University of Padova, Department of Physics and Astronomy
  ``Galileo Galilei'', Vicolo dell'Osservatorio 3, 35122 Padova,
  Italy}} 

\author{
J. J. L{\'o}pez-Moreno
\affil{Instituto  de Astrof\'{i}sica de Andaluc\'{i}a (CSIC), c/
  Glorieta de la Astronomia s/n, 18008 Granada, Spain}} 

\author{
F. Marzari
\affil{ University of Padova, Department of Physics and Astronomy ``Galileo Galilei'', Via Marzolo 8, 35131 Padova, Italy}}

\and

\author{
X. Shi
\affil{Max Planck Institute for Solar System Research,
  Justus-von-Liebig-Weg 3, 37077 G{\"o}ttingen, Germany}}

\clearpage

\begin{abstract}

The phase function of the dust coma of comet 67P has been determined from
Rosetta/OSIRIS images \citep{Bertini17}. This function show a deep minimum at
phase angles near 100$^\circ$, and a strong backscattering
enhancement. These two properties cannot be reproduced by regular models
of cometary dust, most of them based on wavelength-sized and 
randomly-oriented aggregate 
particles. We show, however, that an ensamble of oriented
  elongated particles of a wide variety of aspect ratios, with radii 
$r \gtrsim$10 $\mu$m, and  whose long axes are perpendicular to the
direction of the solar radiation,  
are capable of reproducing the observed phase function. These
particles must be absorbing, with an imaginary part of the refractive index
of about 0.1 to match the expected geometric 
albedo, and with porosity in the 60-70\% range. 
        
\end{abstract}


\section{Introduction}

One of the goals of the Rosetta mission was the characterization of the dust
environment of comet 67P, with several instruments being devoted to this
task. In particular, the  dust phase function has been measured in the
optical range from images taken by the OSIRIS 
cameras \citep{Bertini17}. The phase function was retrieved in a
wide range of phase angles in a 
time interval of about two hours. As these conditions cannot be met from the
ground, the OSIRIS measurements are   
unique.  So far, only for a handful of short-period comets observed from
Earth the phase function is available, albeit being most of the times
restricted to the 
backscattering domain \citep[see][and references 
  therein]{Bertini17}. In addition, for Earth based observations the
phase angle dependence is always inherently mixed with temporal and
heliocentric distance variability in the coma, so that it is often difficult to
disentangle the intrinsic activity from the phase 
effect. Notwithstanding this, an overall agreement of the OSIRIS phase
functions with those derived from the ground is found \citep{Bertini17}.

\section{The OSIRIS phase function}

\cite{Bertini17} have retrieved phase functions from OSIRIS WAC and
NAC images, at varying
heliocentric and nucleocentric distances, with the 
Orange F22 filter (effective wavelength $\lambda_{e\!f\!f}$=642.2 nm), and the 
Green F21 filter ($\lambda_{e\!f\!f}$=537.2 nm) (see their 
figure 4).  During the phase function measurements, the phase angle
  Sun--comet-–spacecraft (i.e., the nucleus elongation) remained
  close to 90$^\circ$.  The geometry of the observations
  can be seen in Figure 1b of \cite{Bertini17}.

For the purpose of comparison with models, we only consider the 
MTP020/071 phase function (obtained near perihelion), which is representative
of most phase curves, although we also display in most 
figures the MTP025/092 curve (r$_h$=2.18 au post-perihelion) for
completeness. That phase curve is the one that 
shows the shallowest slope in the backscattering regime    
(see Figure 1). The corresponding spacecraft (S/C) ranges were   
420 km and 80 km, respectively, indicating that the particles along
the line of sight of the OSIRIS cameras  
were well outside the gas acceleration region \citep[at $\sim$12
  km, see][]{Gerig18}. The large
nucleocentric distances of the S/C (particularly during the perihelion
measurements) and the fact that the pointing is directly away from
the nucleus position would suggest that the phase function can only be
minimally 
affected by different optical depth along distinct lines of sight. In
any case, the true 3D+t nature of the dust distribution, which would
be needed to obtain the optical depth along any line of sight, is unknown, and
will only be unveiled by complex coupled gas and dust dynamics modeling
\citep[e.g.][and references therein]{Crifo06,Zakharov18}.

\section{Modeling the phase function}
 
\subsection{Models assuming size distributions of randomly-oriented
  particles}

All previous models of cometary dust
\citep[e.g.][]{Kimura03,Bertini07,Moreno07,Lasue09,Kolokolova15,Zubko16} 
are built under the hypothesis that the coma is a cloud of particles
in random orientations. However, none of
those models is able to reproduce the phase function curve derived from the
OSIRIS images. In addition, the models based on fluffy,
wavelength-sized, aggregate 
particles \citep[e.g.][]{Kimura03} typically display
 a maximum in the degree of linear polarization that is too high in
 comparison with observations.  \cite{Kimura06} and \cite{KolokolovaMackowski12}
  explain this fact as a computer limitation associated to the limited
  amount of monomers, showing that the maximum of polarization
  decreases as the amount of monomers increase. 
 A rough spheroid model has been
introduced by \cite{Kolokolova15}, which consists of a wide size
distribution of such spheroids. That model matches much better the
observed degree of linear polarization and the color properties, 
but, as it has been stated, the modeled phase functions 
show only a modest backscattering enhancement, and have 
minima at phase angles much smaller than 100$^\circ$,  otherwise in agreement
with the composite phase function curve obtained from a compilation of
ground-based observations from various comets by D. Schleicher
(http://asteroid.lowell.edu/comet/dustphase.html). On the other hand, 
to explain accurately the observations of linear polarization versus
phase angle, models of mixtures of compact and aggregate particles
have been introduced by 
\cite[e.g.][]{Lasue09,Das11}, as well as mixtures of weakly and
highly absorbing agglomerated debris particles \citep{Zubko16}.  

All of those previous models assume either monodisperse distributions
of wavelength-sized particles, 
or polydisperse distributions peaking in the submicrometer range. However,
there are indications that the dominant scatterers in 67P might be 
larger, as indicated by the small amount of submicron and micron-sized
particles detected by Rosetta/MIDAS, much less than initially expected
\citep{Mannel17, Guettler18}.  Also, the analysis of thermal
spectra of the quiescent coma performed with Rosetta/VIRTIS-H 
implies a minimum radius of compact particles of 10 $\mu$m (in
combination with 25\% of fluffy aggregate particles 
in number, whose contribution to the
scattered light is minimal) \citep{Bockelee17a,Bockelee17b}. The study
of the evolution of the 67P dust size distribution by \cite{Fulle16b}
confirms that the coma brightness at perihelion and later is in fact dominated
by particles $<$100 $\mu$m. In addition, the
analysis of a large ground-based 
image dataset by Monte Carlo dust tail modeling \citep{Moreno17}
agrees with that constraint. This model replicated tail images from
the current apparition, from about 4.5 au pre-perihelion to 3 au
post-perihelion, as well as trail data from the current and previous
orbits. The minimum model particle size was 10 $\mu$m in all 
cases. 

Given these facts, we started by trying models having randomly-oriented
particles of sizes larger than the nominal incident wavelength, assumed at 
$\lambda$=0.6 $\mu$m. For this task, 
we used available light scattering codes such as the T-matrix code of spheroids
\citep{Mishchenko96}, the Multi-Sphere T-matrix (MSTM) code \citep{Mackowski11},
and the geometric optics code by \cite{Macke96} for ellipsoids. Except for
the geometric 
optics code, the computing time and memory requirements are a rapidly 
increasing functions of the size parameter, $X$=2$\pi r/\lambda$. This
imposes strong limits to the 
simulations, for which we should keep the number of possible
combinations of input parameters to a minimum. 

One important constraint to the model is the observed geometric albedo. The
nucleus geometric albedo at 649 nm is 0.0677$\pm$0.0039
\citep{Fornasier15}, and it is reasonable to assume that large particles
in the coma will 
have the same geometric albedo, provided they do not experience any change
in their physical properties after ejection, such as sublimation of
volatiles or fragmentation. These processes have been found to be
negligible in the Rosetta studies of the 67P's coma \citep{Fulle15}. 

To meet the geometric albedo constraint we must assume 
absorbing particles. The precise value of the mean refractive index
of 67P dust coma particles is unknown. We used a refractive index of
$m$=1.6+0.1$i$, a value comprised between low-absorbing silicates 
and strongly absorbing organic and carbonaceous materials 
at red wavelengths \citep[see e.g.][]{Jenniskens93}. To
calculate the scattering pattern of relatively large particles, we
used the MSTM code for an array of spherules. Since we are searching
for scatterers giving a high backscattering enhancement, we started by
following a procedure similar to that devised by \cite{Mishchenko07}. To
demonstrate the 
onset and development of the coherent backscattering mechanism, which
is the responsible for the 
backscattering enhancement, \cite{Mishchenko07} considered a certain
scattering volume filled with an increasing number of small
scatterers. We performed a similar calculation by considering a
spherical volume  
of $r$=2 $\mu$m containing 500, 1000, and 2500 randomly placed 0.1-$\mu$m 
radius spherules, giving a porosity, $P$, of 93.7\%, 87.5\%, and
69.0\%, respectively (see Figure 1, top panels). To obtain the
scattering functions, we assumed that 
the array of scatterers is illuminated from a given direction,  
and then compute the azimuthally-averaged 
scattering matrix, i.e., the average scattering matrix for all the
scattering planes about the direction of the incoming beam. Then, for
a given number of spherules, we repeated 
the simulation for a large number of such randomly-generated
targets. This procedure 
is much less CPU time consuming than generating the scattering matrix
for a single 
randomly-oriented scatterer. 

The resulting phase functions are 
displayed in Figure 1. Also shown are the results corresponding to a
single homogeneous $r$=2 $\mu$m sphere 
calculated from Mie theory, just to show the large discrepancies of that
simple model with the MSTM models in the phase function. 
Concerning the MSTM models, we see that as the number of spherules in the
volume increases, the backscattering enhancement also increases. In
order to check whether this trend is 
maintained for still lower porosities, we first increase the number of
spheres up to $\sim$3100, corresponding to $P$=61\%, but we obtained
very similar results to that of $P$=69\%. This $P$=61\% is essentially 
the lower limit reachable with our simple random packing procedure. To
decrease $P$, we had to use a more complex 
technique than random packing. Among the techniques available, 
we used a ``falling down'' algorithm \citep{FulleBlum17}. With 
this code, we were able to generate spherical volumes having $P$ in the range
55\%-45\%, i.e., containing some 3600 to 4400 spherules. At these
porosity levels, we noticed a reversal in the trend, i.e., a decrease in
backscattering enhancement as compared with that found in the
$P\sim$60-70\% range. We then conclude that this porosity range is the one that
gives the highest backscattering enhancement. We underline that this
conclusion is based on the specific spherical volume described above
(see Figure 1, top panels). Larger volumes would have been  
desirable to test, but, as stated above, we are strongly constrained
by computer memory and CPU time limitations.  

It is interesting to note that the porosity range for which the
highest backscattering enhancement is observed agrees with the nucleus
porosity, estimated at 71$\pm$8\% by \cite{Fulle16a}, and is also consistent
with the value found specifically for the upper layers of the
surface, which is below 74\% from Philae SESAME/CASSE and MUPUS
\citep{Knapmeyer18}. It is also not far from the 
Rosetta/MIDAS and Rosetta/GIADA results for relatively compact
particles \citep[$\sim$50\%, see][]{FulleBlum17}.   

The following step was to explore other configurations of
arrays of spherules of total size larger than the wavelength 
to test whether an improvement in the fits could be   
achieved. We built larger particles by increasing the number of
spherical monomers, but owing to the limitations in computer time
and memory, we had to combine larger spherical monomers with smaller
ones, although we could not find a satisfactory fit in any case. An example of
those configurations is shown in Figure 2, in which   
a cluster of four 2-$\mu$m spheres is randomly peppered  
by a total of 2000 0.1 to 0.15 $\mu$m radius 
spherules having a total equivalent spherical radius (radius of a
sphere of equal 
volume), $r_{eq}\sim$3.4 $\mu$m. This particular arrangement gives an
overall better 
fit to the phase  
function, mainly for the forward lobe, but fails at reproducing the
observed backscattering enhancement.         

To test the performance of still larger particles, we had to resort to
the geometric optics code. As mentioned, we used the code 
for ellipsoidal particles by \cite{Macke96}. We have
tested a variety of ellipsoids of different sizes and axes ratios, but
the computed phase functions are all flat at backscattering, similarly
to large homogeneous spheres. In
conclusion, none of the considered models of randomly-oriented particles at a 
wide range of sizes can reproduce reliably the OSIRIS phase function.

\subsection{Models assuming size distributions of particles aligned to
the solar radiation direction}
    
The presence of aligned particles in the cometary environment has been
subject of research for many years, since the early work of
\cite{Dolginov76}. In the nucleus acceleration region, comet gravity,
and aerodynamic forces should dominate 
\citep{Fulle15,Ivanovski17a}. Outside that region, radiative torques,
although ignored 
for many years, might play an important role \citep[see
  e.g.][]{Draine96,Lazarian03,Rosenbush07}. Interestingly, alignment
of the particles with respect  
to the solar radiation will place the long axes of the particles 
perpendicular to radiation \citep{Rosenbush07}.

For the radiative torques to be efficient as the alignment mechanism, the
  characteristic time of alignment should be shorter than the flight
  time of the particles along the line of sight of the OSIRIS
  camera. \cite{Kolokolova16}, based on a study by 
  \cite{HoangLazarian14}, estimate that for the 67P environment, a 10-$\mu$m
  oblate spheroid would be aligned in $\sim$3$\times$10$^5$ s. Assuming
  a radial trajectory and a constant speed of 10 m s$^{-1}$
    \citep{Lin16},  that particle would become aligned at a distance
    of 3000 km from the nucleus. Thus, it appears that this mechanism
    alone would not be 
  sufficient to provide the required alignment, at least for those
  particles in the vicinity of the S/C. However, in addition
  to radiative torques, there is another mechanism that could be playing a
  role, which is the mechanical alignment by gas-particle relative
  motion \citep{Gold52} and aerodynamic force 
  \citep[e.g.][]{Ivanovski17b}. The Gold alignment mechanism,  
which is efficient for supersonic gas flow 
\citep[as in a cometary coma, see e.g.][]{Zakharov18}     
 would be characterized by particles having their longer 
axes  directed  preferentially  along  the  gas flow 
due to the tendency of the particle to minimize its cross section
in the gas flow \citep{Rosenbush07}. The gas flow is radial at
distances greater than about 20 km from the nucleus
\citep{Marschall16}. The particles experience this radial gas flow
until the gas decoupling distance,
at some 20 nuclear radii (i.e., about 30 km).  After that, there is no
other mechanism 
that can affect their orientation except radiative torque which 
produce alignment in the same direction - long axis perpendicular to
the radiation, thus, keeping the original gas alignment.  This
direction is  
nearly perpendicular to the direction of the solar radiation for particles in
the vicinity of the S/C during the perihelion phase
function measurements \citep[see Figure 1b by][]{Bertini17}. Thus, the
radiative torque and the Gold mechanism working together might
explain the alignment of the particles. However, although a quantitative 
evaluation of the efficiency of 
those mechanisms in terms of the physical properties of
the particles should be certainly performed, this is beyond the scope
of this paper. 
    
To model aligned scatterers, we consider that the particles 
have their long axes perpendicular to radiation, and calculate the
azimuthally averaged scattering matrix. For these
simulations, we assumed spheroidal particles, because in this way we
can encompass a wide range of sizes with same geometry. The spheroids
have axes (a,b,c), where a=b$>$c for oblate, and a$>$b=c for prolate
spheroids, respectively. The axis ratio is defined as
$\epsilon$=a/c. Thus, oblate 
spheroids are set with the c-axis parallel to the solar radiation, and prolate
spheroids are set with the a-axis perpendicular to the solar radiation. 

Although we know that the dominant scatterers in 67P must be larger than
the wavelength, for completeness, we performed simulations for a wider
range of sizes, starting from the submicron domain. This will show that
the dominant scatterers must be, in fact, much larger. Thus, for 
distributions of small sized particles, 
we used the T-matrix method for spheroids by 
\cite{Mishchenko96}. We start by considering oblate spheroids only, because 
they provide a higher backscattering enhancement than prolate
spheroids for a given $r_{eq}$. In addition, oblate spheroids might indeed
  provide a good description of cometary dust particles as has been
  found from laboratory experiments by \cite{StephensGustafson91}, and
  recently by \cite{Bischoff18}. Both studies dealt with mixtures of
  volatile and refractory materials and showed that during the
  sublimation of volatile materials, the refractory materials form a
  thin mantle that eventually cracks, being lifted from the
  surface. Bischoff et al. show that these cracked pieces have a
  diameter of up to 5 to 10 times larger than the thickness of the mantle,
  i.e. they are flake-like particles resembling oblate spheroids of large axis
  ratio.   

We generated scattering matrices
 for size distributions of such spheroidal particles, for which we assumed
 $\epsilon$=2. Figure 3 shows two 
 simulations with such spheroids being distributed according to a
 differential power law of
 index --3.5 (a typical value for 67P dust coma), and with three 
 different values of  the minimum size, 0.1, 0.3, and 1.5 $\mu$m, the
 maximum size being r$_{max}$=2.5 $\mu$m in all cases. From Figure
 3,  it  is clear that the larger the sizes the better the agreement 
 with the OSIRIS phase function. On the other end, submicron-dominated
 size distribution functions fail at fitting the phase
 function. 

Sizes substantially larger than the maximum radius assumed in the previous
model of spheroids (r$_{max}$=2.5 $\mu$m with $\epsilon$=2) cannot be 
efficiently handled with the T-matrix code. To build larger particles, 
we fill oblate spheroidal volumes with spherules
of the order of the nominal wavelength or smaller, as input to the MSTM
code, with an average porosity of 65-70\%, i.e., near the optimal values
found for the 2$\mu$m spherical volume described in the previous 
section. We built oblate spheroidal volumes with axial ratios in the
range $\epsilon$=2-3, and with equivalent spherical radius of 3.1 $\mu$m, 4.6
$\mu$m, and 7.0  $\mu$m, filled with approximately 1000, 1500, and
4000 spherules, respectively, where the radii of the spherules 
range from 0.1 to 0.6$\mu$m. Owing to computing time
limitations, the number of realizations for each size was limited to
256, 36, and 20, respectively. Figure 4 shows the resulting phase
functions for the three 
sizes, where we can notice again how the agreement with the typical 
OSIRIS phase function tends to improve as the particle size increases,
in both forward 
and backward lobes. 

So far, only oblate spheroids up to a size of $r_{eq}$=7 $\mu$m have
been considered as model particles. We have to extend the upper size limit
in the model to larger sizes, as measured by several
instruments on board Rosetta, and inferred from the
ground. In addition, we have to extend the particle shape distribution
by taking into account a mixture of oblate and prolate particles, as would be
required to simulate any natural dust sample. Thus, for a given equivalent
spherical radius,  we assumed a uniform distribution of axes  
ratios, i.e., the same amount of prolate and oblate spheroids. 

To generate the shape-averaged scattering matrix, we used the 
individual matrices calculated using the geometric optics code by
\cite{Macke96}. Figure 5 shows a comparison of the results obtained
for three equivalent spherical radii ($r_{eq}$=10 $\mu$m, $r_{eq}$=100
$\mu$m, and $r_{eq}$=1000 $\mu$m) with the OSIRIS phase function, for
a shape distribution having axes ratios in the interval 
$\epsilon$=[0.25,4].  This wide axes ratio distribution is not
surprising,  in view of the findings by \cite{Fulle17} who inferred  
aspect ratios of 5 or 
more in order to explain the bulk densities of the particles measured by
GIADA during the whole mission.   

As it is seen, the synthetic phase functions
are close at backscattering, but the one for the smaller radius 
gives a better fit at large phase angles. In addition, the resulting
geometric albedo  
is 0.06, very close to that obtained at the nucleus surface
\citep{Fornasier15}. The inferred particle size is consistent with the
ground-based tail modeling, where differential power-law size
distributions of index 
$\lesssim$--3 and  minimum particle radius of 
10 $\mu$m were found to be consistent with the observations 
\citep{Moreno17}. This is also consistent with MIDAS results, where a
very small 
amount of micron-sized particles or smaller were detected
\citep{Mannel17,Guettler18}, and with VIRTIS-H thermal spectra
modeling where minimum sizes of relatively compact 10 $\mu$m particles
(combined with 25\% fractals in number) following a size distribution of
power index --3 or smaller are found 
\citep{Bockelee17a,Bockelee17b}. It is also important to note that 
the phase function shape is maintained for up to mm-sized particles in
the backscattering lobe, explaining in that way the results obtained
by \cite{Fulle18}, in that the OSIRIS phase function should be applied
to dust diameters up to 2.5 mm at least for the conversion of dust
brightness into dust sizes.

\section{Conclusions}

We have shown that regular models of cometary dust based on
wavelength-sized and randomly-oriented aggregate 
particles cannot reproduce the dust phase function determined from
Rosetta/OSIRIS observations. Such phase function can, however, be 
adequately fitted by assuming that the main scatterers in 67P coma are
large compared to the visual wavelengths 
($\gtrsim$10 $\mu$m), show a wide distribution of aspect ratios, and
are aligned with their long axes perpendicular 
to the solar radiation direction. Porosity levels for the particles 
of 60-70\%  would be particularly 
favoured, since those particles would give the largest backscattering
enhancement. For a refractive index of $m$=1.6+0.1$i$, the
geometric albedo for such particles would be $\sim$0.06, i.e., close
to that measured for the nucleus surface. 

\acknowledgments

OSIRIS  was  built  by  a  consortium  of  the  Max-Planck-Institut 
f\"ur Sonnensystemforschung, in G\"ottingen, Germany, CISAS-University
of Padova, Italy, the Laboratoire d’Astrophysique de Marseille,
France, the Instituto  de  Astrof\'\i sica  de  Andaluc\'\i a,  CSIC,
Granada,  Spain,  the  Research  and 
Scientific Support Department of the European Space Agency, Noordwijk, The
Netherlands,  the  Instituto  Nacional  de  T\'ecnica  Aeroespacial,
Madrid,  Spain, the  Universidad  Polit\'ecnica  de  Madrid,  Spain,
the  Department of  Physics 
and Astronomy of Uppsala University, Sweden, and the Institut f\"ur 
Datentechnik und Kommunikationsnetze der Technischen Universit\"at
Braunschweig, Germany. The support of the national funding agencies of
Germany (DLR), France 
(CNES), Italy (ASI), Spain (MEC), Sweden (SNSB; Grant No. 74/10:2), 
and the  ESA  Technical  Directorate  is  gratefully  acknowledged.  H. 
Rickman  was also supported by Grant No. 2011/01/B/ST9/05442 of 
the Polish National Science Center. We thank the ESA teams at ESAC,
ESOC and ESTEC for their 
work in support of the Rosetta mission
We thank the Rosetta Science Ground Segment at ESAC, the Rosetta
Mission Operations Centre at ESOC, and the Rosetta Project at ESTEC
for their outstanding work enabling the science return of the Rosetta
Mission.

We thank Evgenij Zubko for his constructive comments on the paper, and
to Daniel Mackowski, Michael Mishchenko, and Andreas Macke for making
their light scattering codes available.  

This work was supported by contracts AYA2015-67152-R and
AYA2015-71975-REDT from the Spanish
Ministerio de Econom\'\i a y Competitividad.

\clearpage

\begin{figure} 
\centering
  \begin{tabular}{ccc}

    \includegraphics[width=.33\textwidth,angle=0]{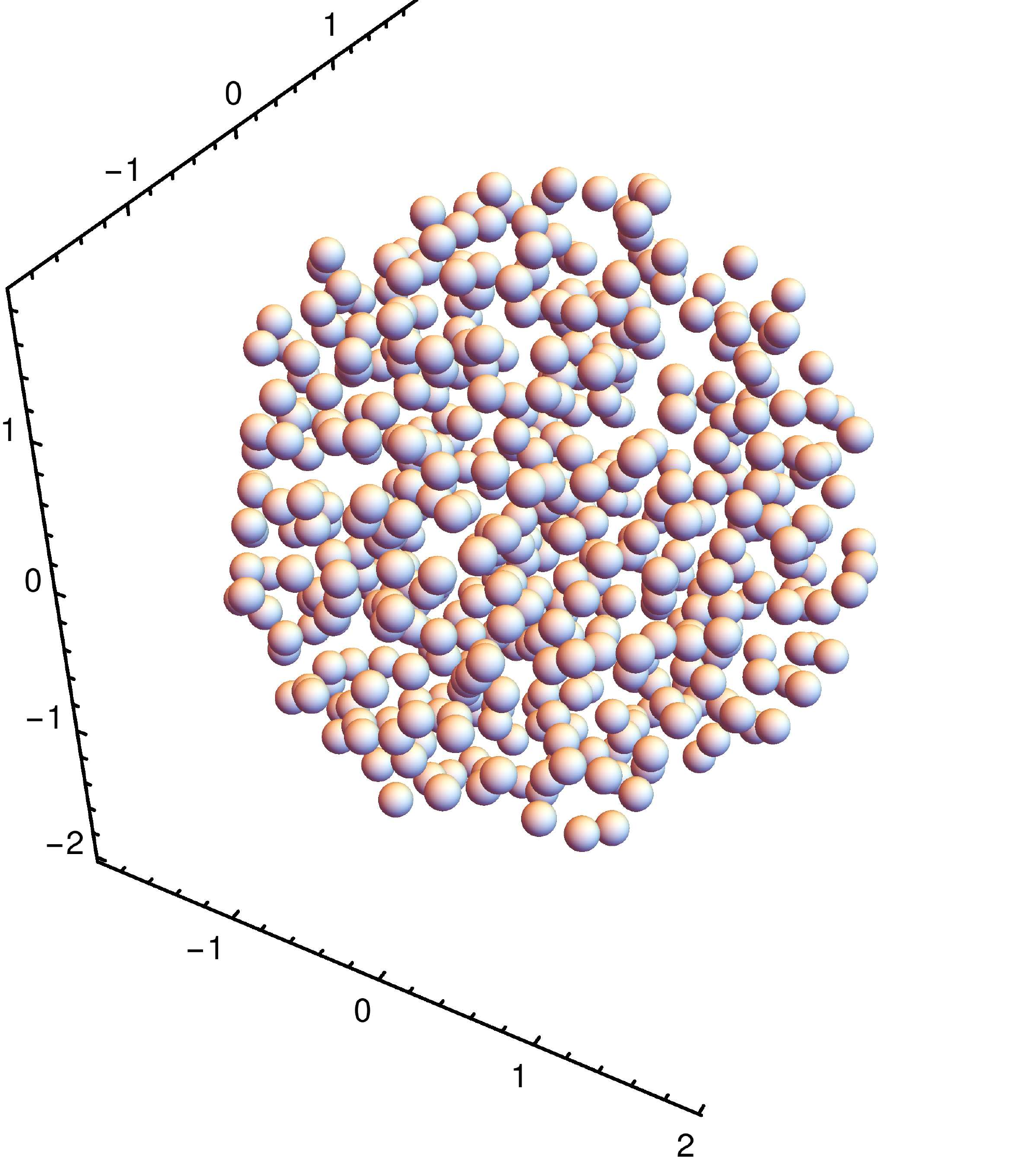} &
    \includegraphics[width=.33\textwidth]{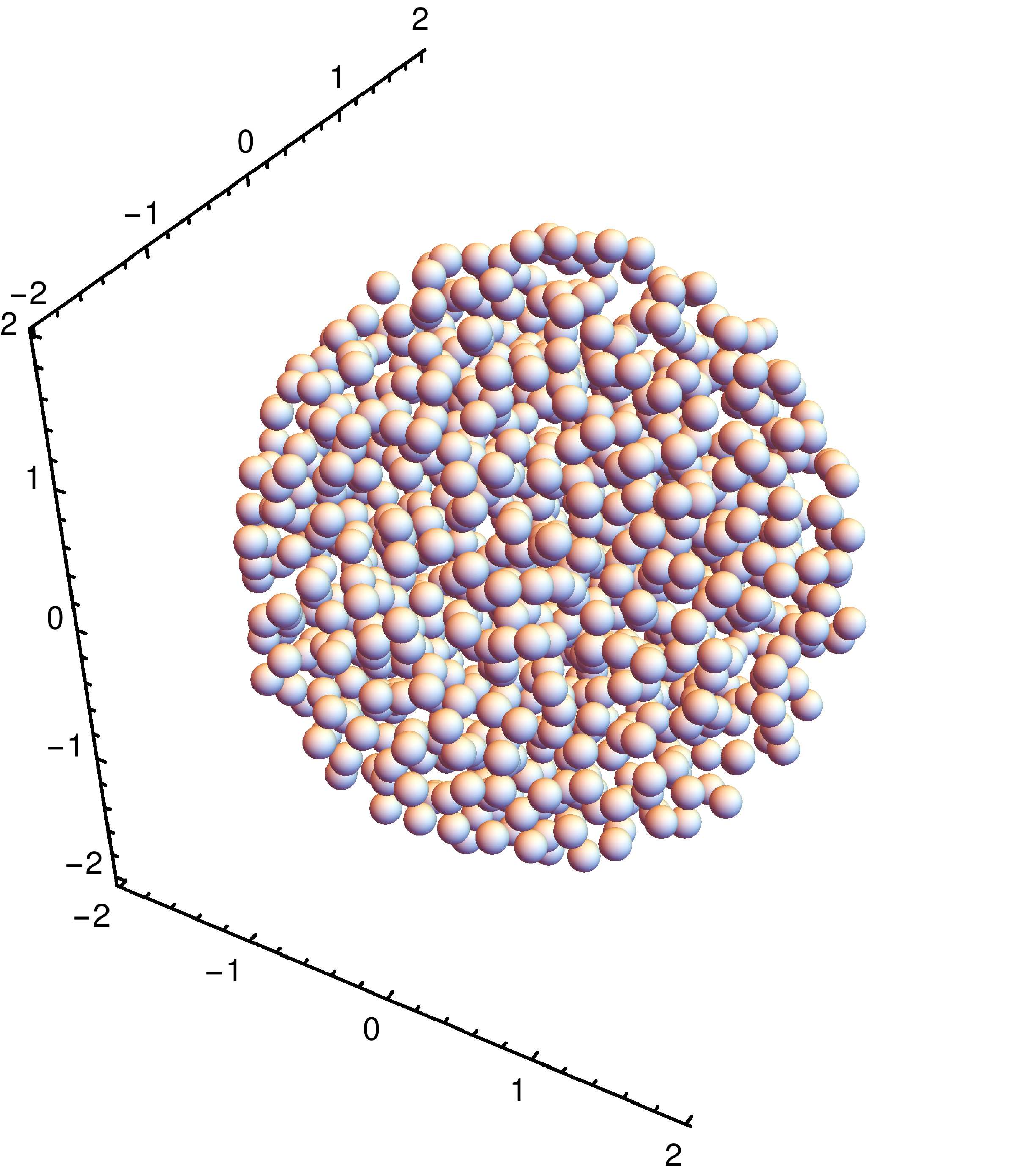} &
    \includegraphics[width=.33\textwidth]{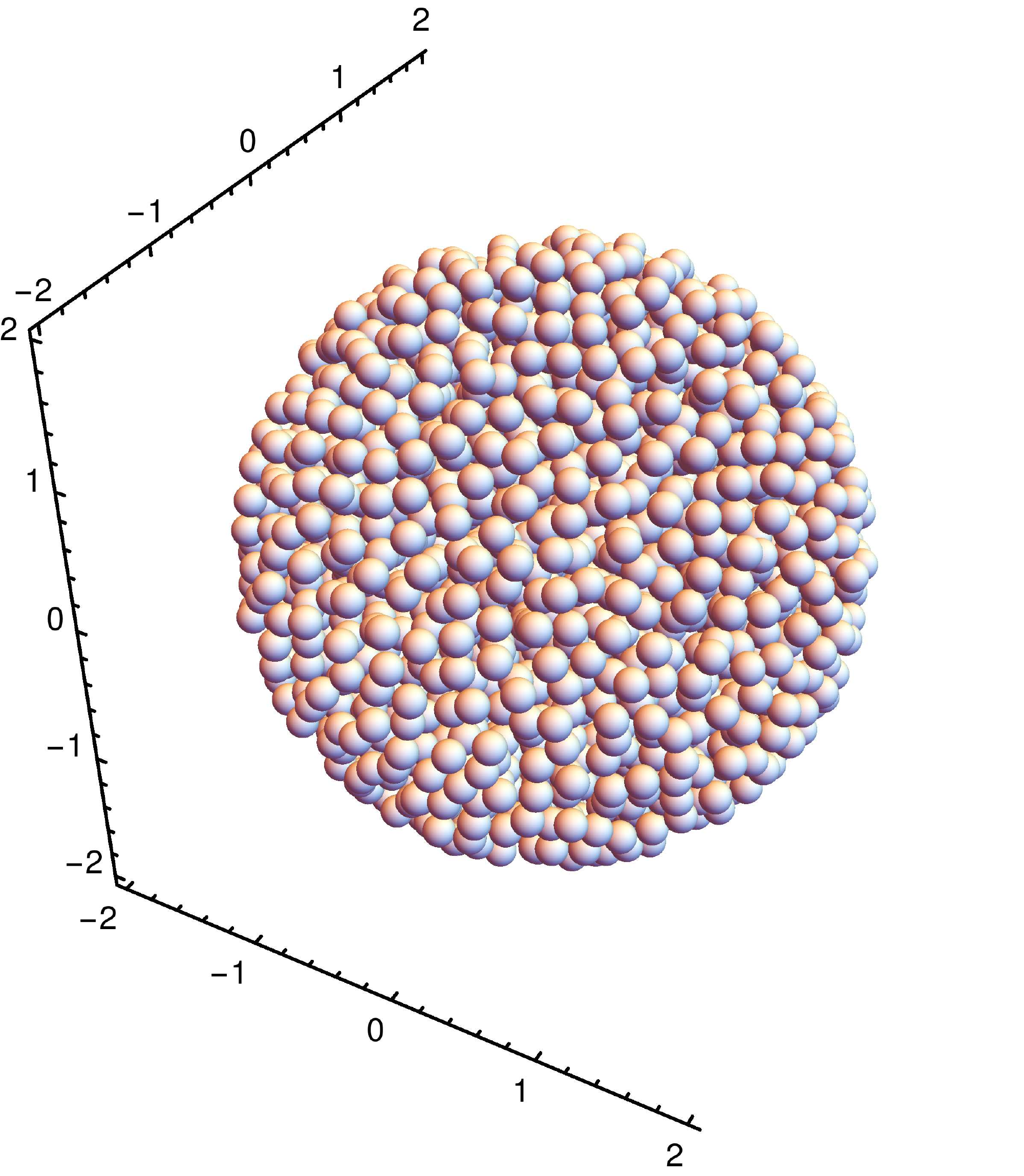} \\

    \multicolumn{3}{c}{\includegraphics[width=.55\textwidth,angle=-90]{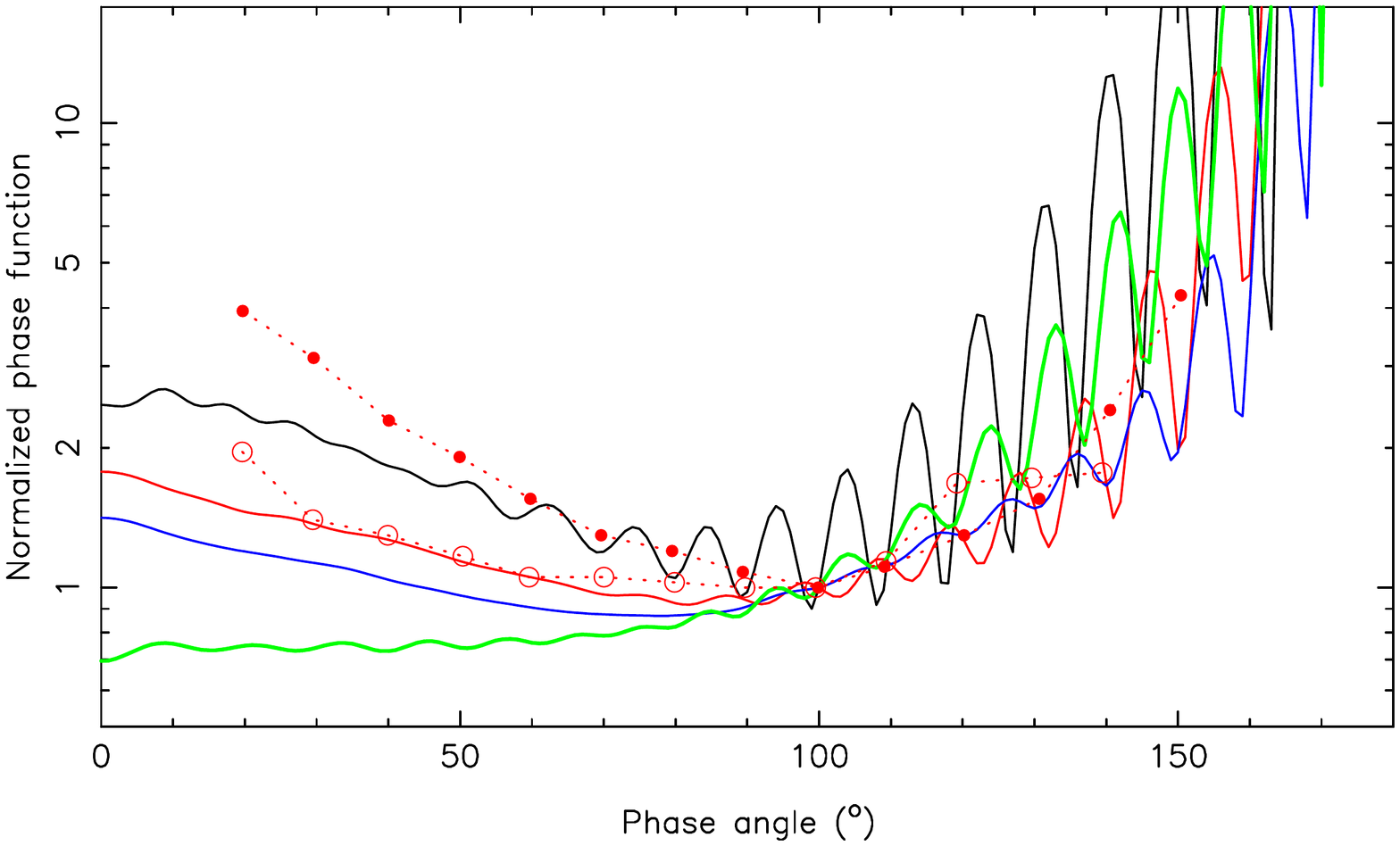}}

\end{tabular}
\caption{Backscattering enhancement test. Top three panels: 2-$\mu$m
  radius spherical volumes filled 
  with (left to right) 500, 1000, and 2500 randomly located 0.1-$\mu$m radius
  spherules, leading to porosities of 93.7\%, 87.5\%, and 69.0\%,
  respectively. XYZ labels are in $\mu$m. 
Bottom panel: OSIRIS phase functions (red solid
  circles for MTP020/071 and red open circles for MTP025/092) compared
with MSTM calculations for the above clusters. Blue, red, and black thin
lines correspond to porosities of 93.7\%, 87.5\%, and 69.0\%, respectively. The
thick green line corresponds to the Mie theory phase function  
for a homogeneous and compact sphere of 2-$\mu$m radius. All phase functions are
normalized to unity at 
a phase angle of 100$^\circ$.}
   \label{fig1}
\end{figure}

\clearpage

\begin{figure} 
\centering
\includegraphics[width=.55\textwidth,angle=-90]{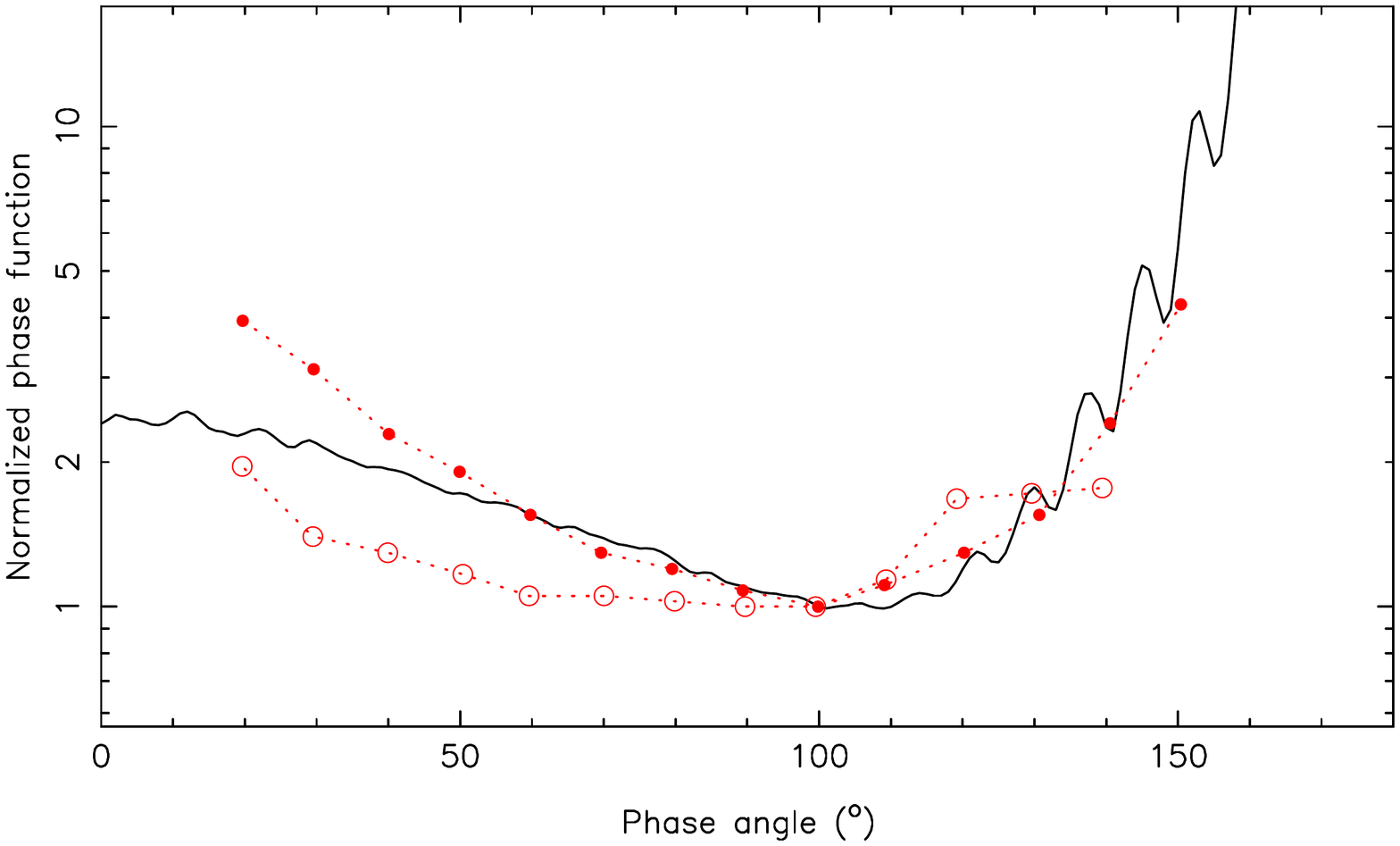}\llap{
\makebox(95,-313)[l]{\includegraphics[height=-3.2cm,angle=180]{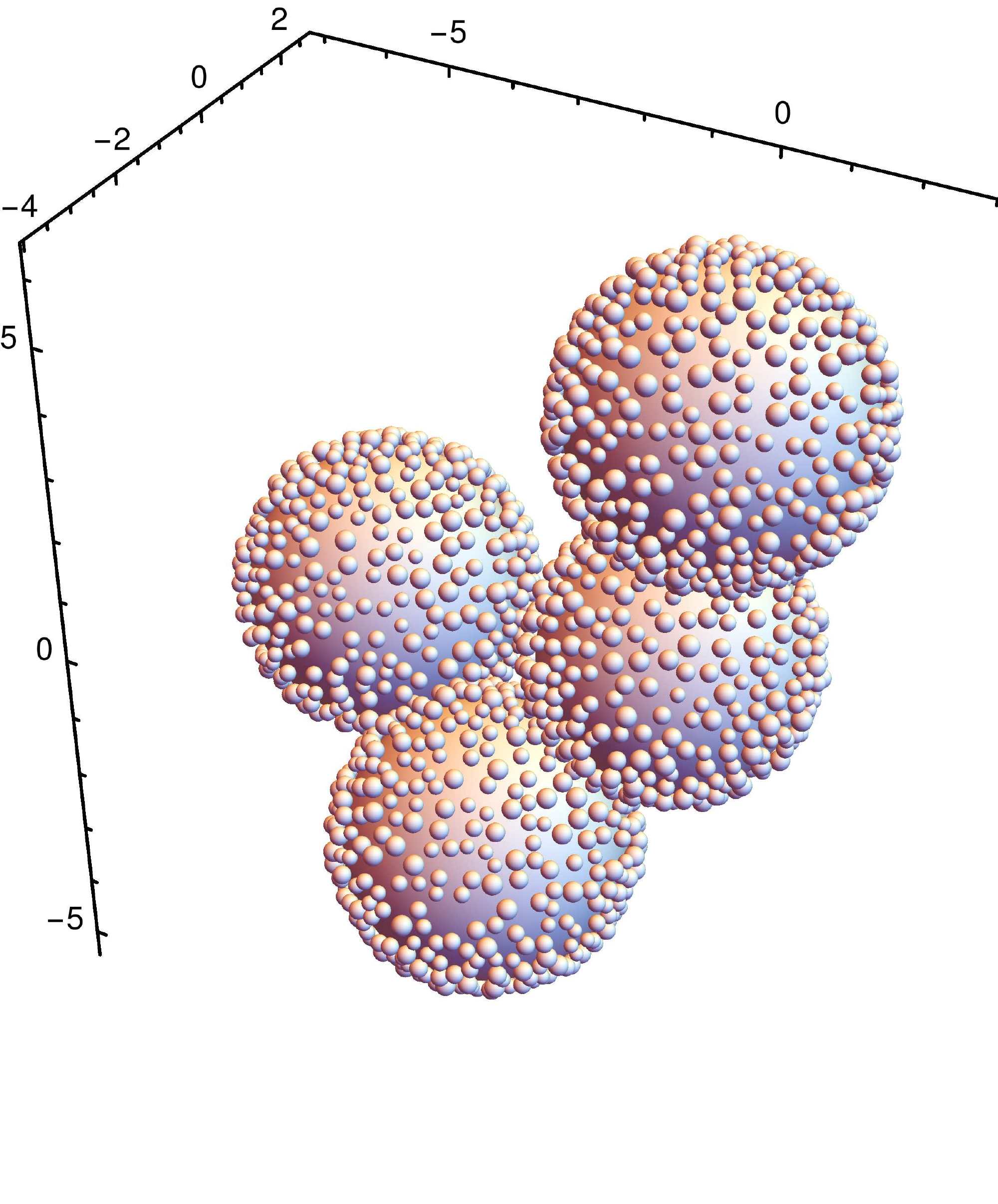}}}
\caption{OSIRIS phase functions (red solid
  circles for MTP020/071 and red open circles for MTP025/092) compared
with MSTM calculations (solid black line) for peppered sphere   
clusters such as that shown on the inset. XYZ
labels on the inset are given in $\mu$m. All phase
functions are normalized to unity at a phase angle of  
100$^\circ$.} 
\label{fig2}
\end{figure}

\clearpage

\begin{figure}
\centering
\includegraphics[width=.55\textwidth,angle=-90]{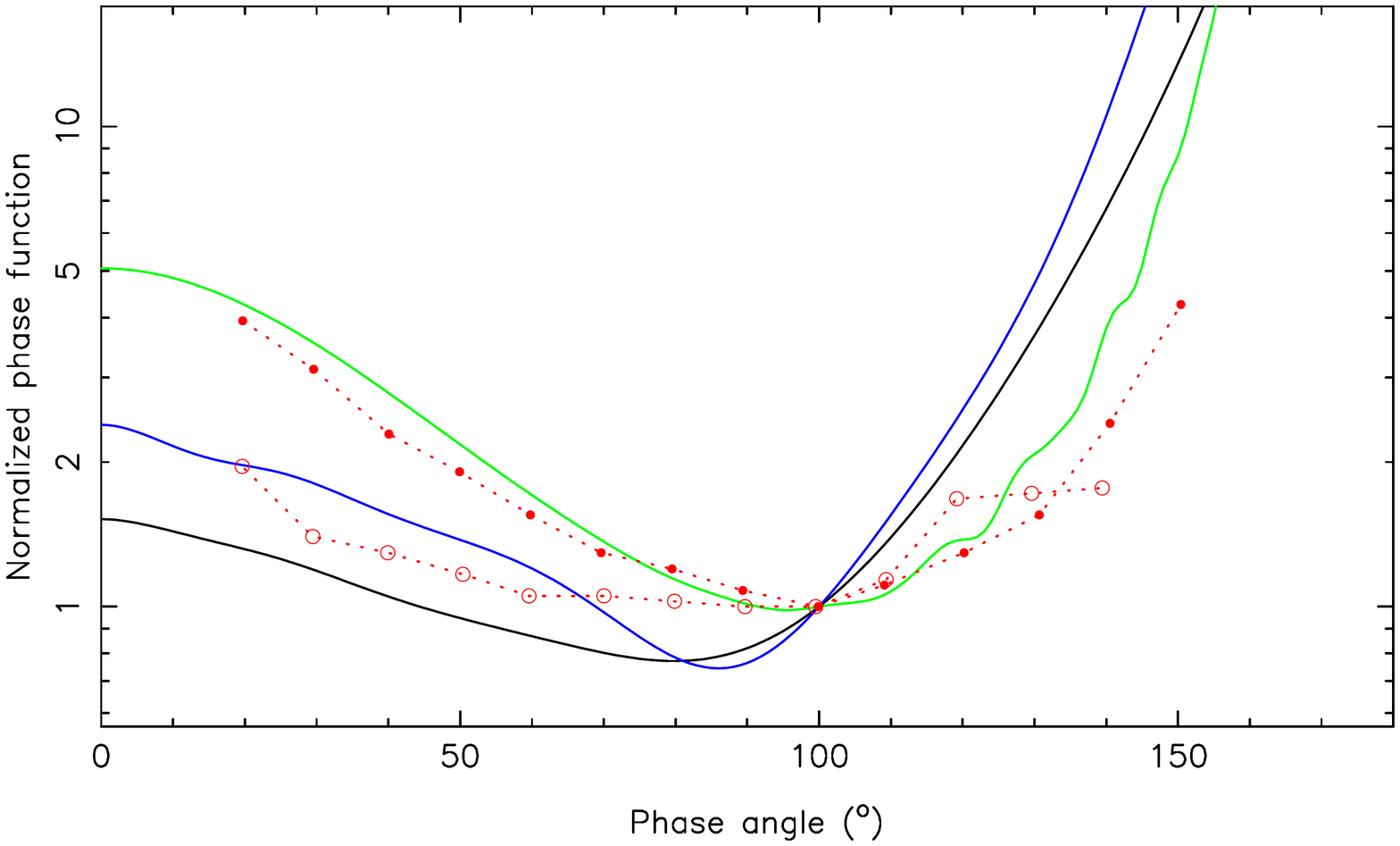}\llap{
\makebox(100,-250)[l]{\includegraphics[height=-2.2cm,angle=180]{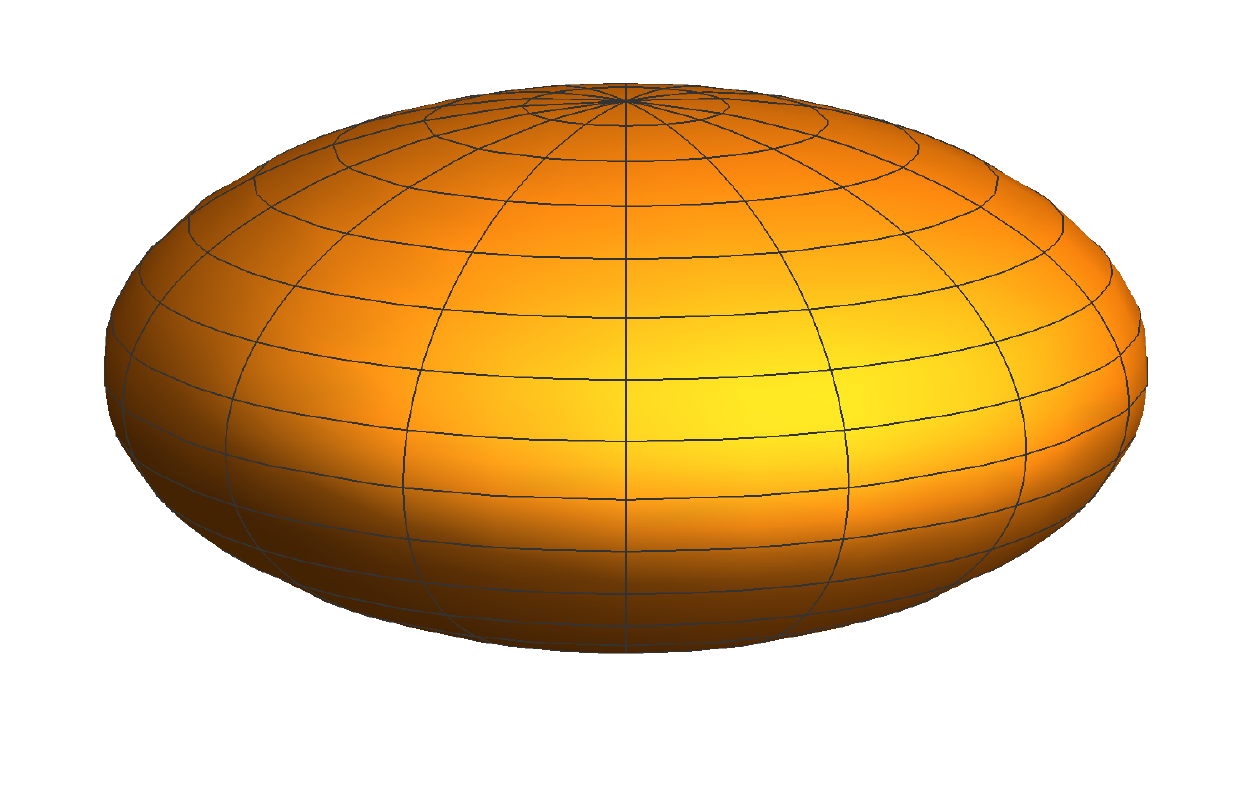}}}
\caption{OSIRIS phase functions (red solid
  circles for MTP020/071 and red open circles for MTP025/092) compared
  with T-matrix calculations for size distributions of oriented, compact, and 
oblate ($\epsilon$=2, as that shown) spheroids: the black,
  blue, and green lines are for size distributions having
  r$_{min}$=0.1, 0.3, and 1.5 $\mu$m, respectively, with r$_{max}$=2.5
  $\mu$m. All phase functions are normalized to unity at a phase angle of 
100$^\circ$.}
   \label{fig3}
\end{figure}

\clearpage

\begin{figure}[ht]
\centerline{\includegraphics[scale=0.8,angle=-90]{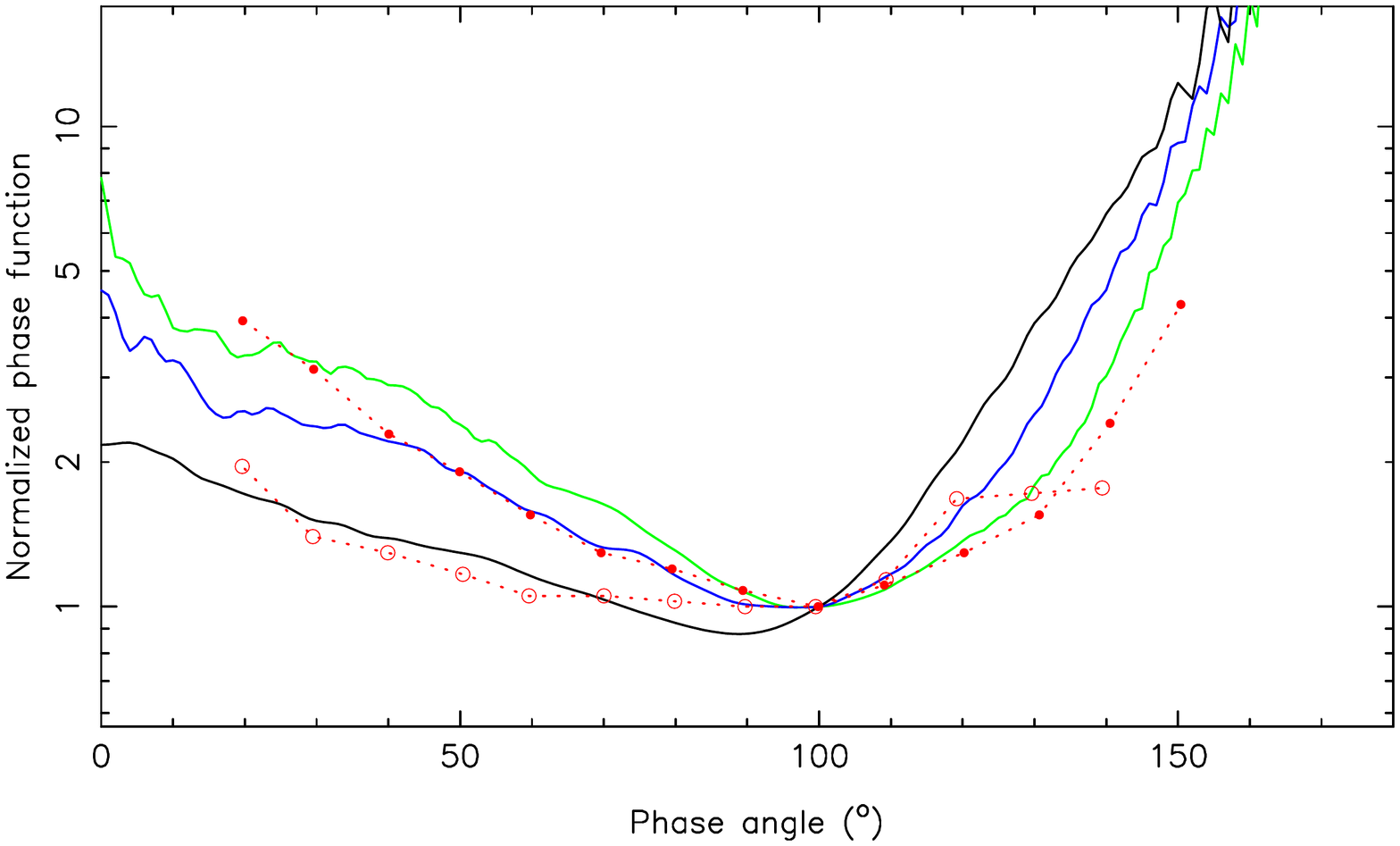}\llap{
\makebox(250,-460)[l]{\includegraphics[height=-3.8cm,angle=180]{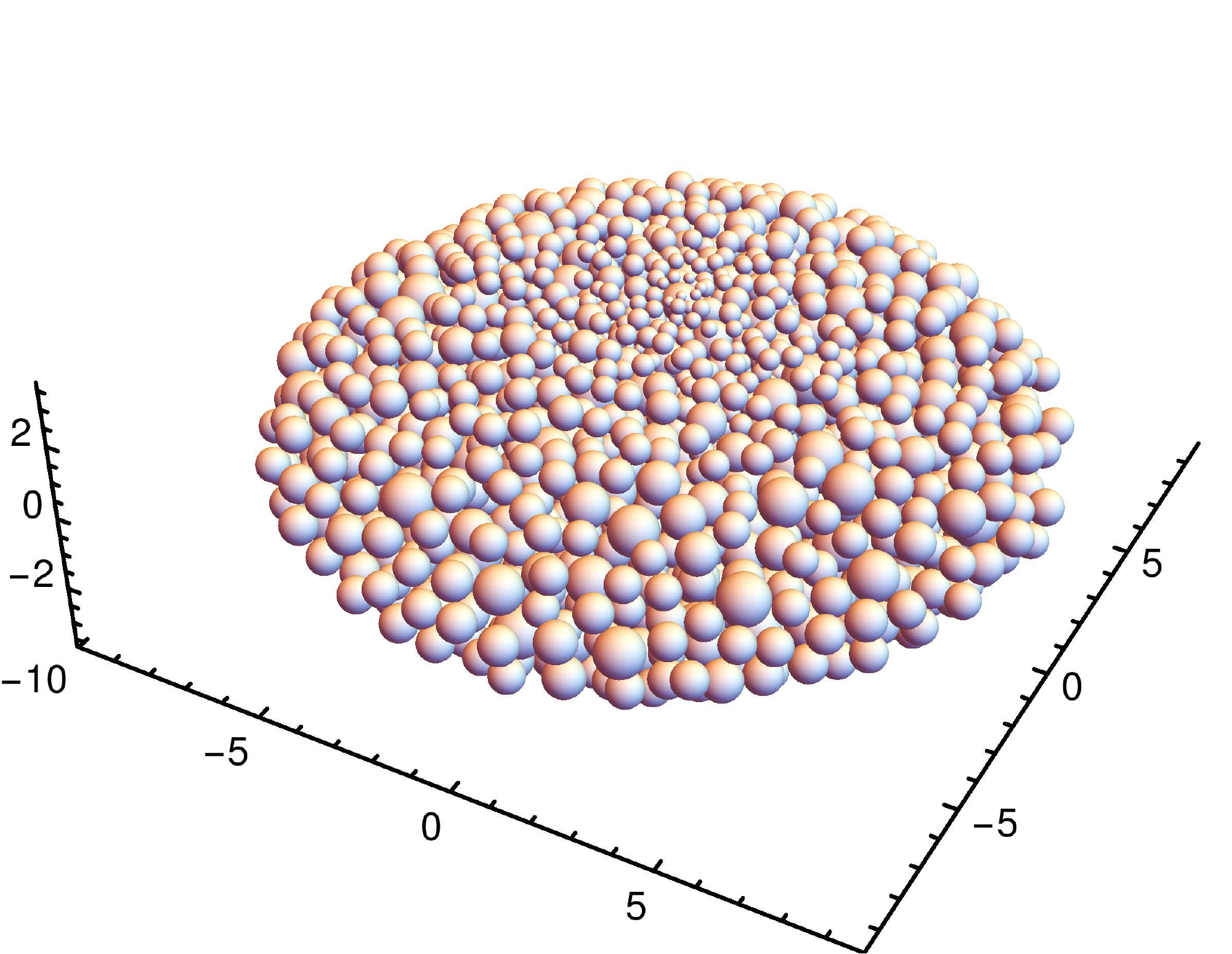}}}}
\caption{OSIRIS phase functions (red solid
  circles for MTP020/071 and red open circles for MTP025/092) 
compared with MSTM simulations of oriented
  oblate spheroidal volumes filled with 0.1 to 0.6-$\mu$m
  spherules. Black, blue, and green lines are for equivalent spherical
  radii of 3.1, 4.6, and 7.0 $\mu$m. The inset illustrates the array
  of spherules for the 
  largest particle, where the XYZ labels are in $\mu$m. All phase
  functions are normalized to unity at a phase angle of 100$^\circ$. }
   \label{fig4}
\end{figure}

\clearpage

\begin{figure}[ht]
\centerline{\includegraphics[scale=0.8,angle=-90]{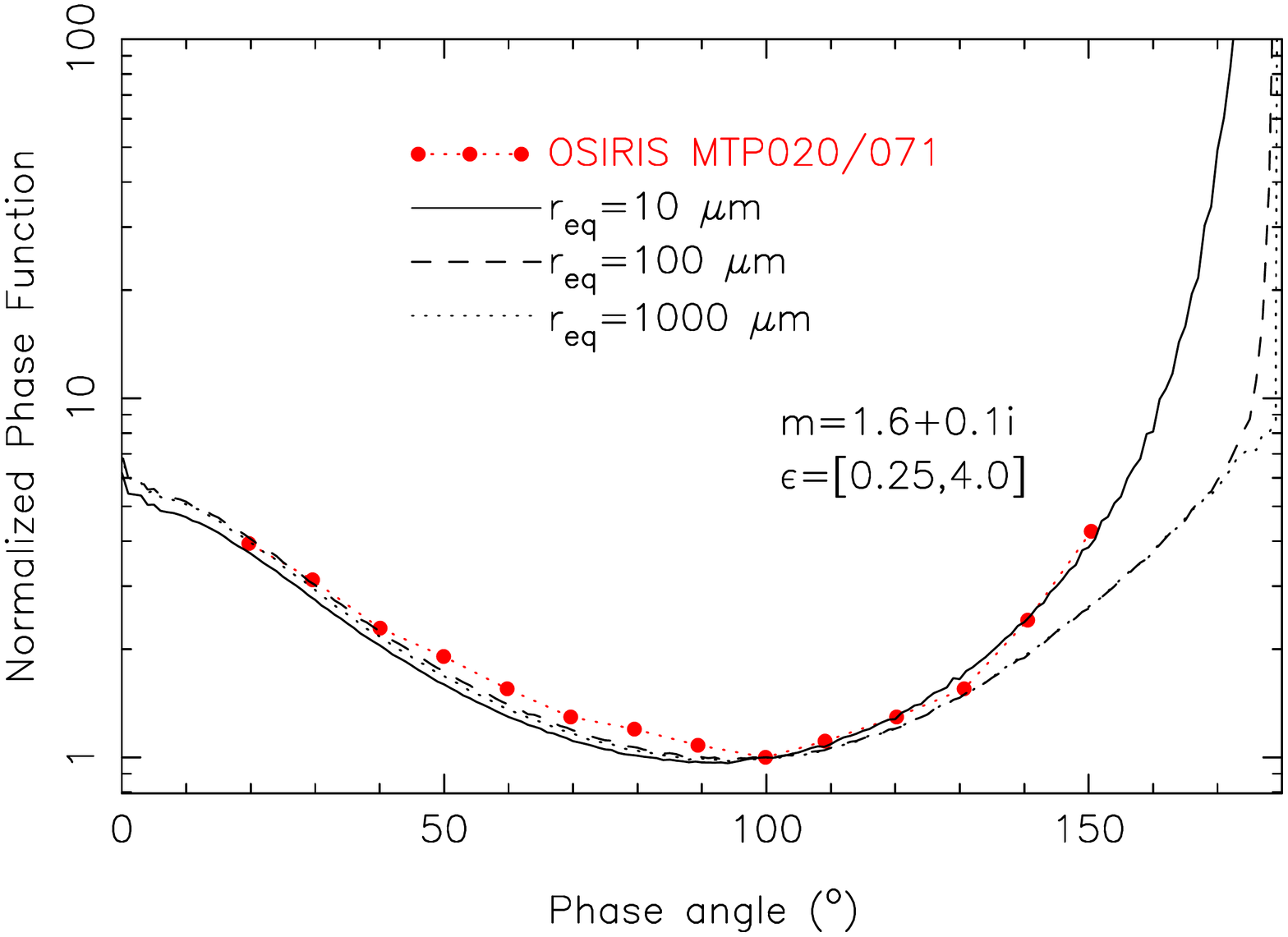}}
\caption{OSIRIS phase function measurements for MTP020/071 (solid red
  circles) compared to geometric optics simulations for 
 three distributions of spheroids with uniform
  axes ratio in the $\epsilon$=[0.25,4.0] interval, and having
  equivalent spherical 
  radii of 10 $\mu$m (black solid line), 100 $\mu$m (black dashed 
line), and 1000 $\mu$m (black dotted line). All phase functions are
normalized to unity at 100$^\circ$ of phase 
  angle. As in all the simulations in
this work, the refractive index is 
set to $m$=1.6+0.1$i$.}
   \label{fig5}
\end{figure}


\begin{thebibliography}{}


\bibitem[Bertini et al.(2007))]{Bertini07} Bertini, I., Thomas, N., \&
  Barbieri, C., 2007, \aap, 461, 351 

\bibitem[Bertini et al.(2017))]{Bertini17} Bertini, I., La Forgia, F.,
  Tubiana, C. et al., 2017, \mnras, 469, 404

\bibitem[Bischoff et al.(2018))]{Bischoff18} Bischoff, D., Gundlach,
  B., Neuhaus, M. et al., 2018, \mnras, submitted

\bibitem[Bockel\'ee-Morvan et al.(2017a)]{Bockelee17a}
  Bockel\'ee-Morvan, D., Rinaldi, G., Erard, S., et al., 2017a, \mnras,
  469, 443

\bibitem[Bockel\'ee-Morvan et al.(2017b)]{Bockelee17b}
  Bockel\'ee-Morvan, D., Rinaldi, G., Erard, S., et al., 2017b, \mnras,
  469, 842

\bibitem[Crifo(2006)]{Crifo06} Crifo, J.F., 2006, Adv. Space Res., 38, 1911

\bibitem[Das et al.(2011)]{Das11} Das, H.S., Paul, D., Suklabaidya,
  A., et al., 2011, \mnras, 416, 94

\bibitem[Dolginov \& Mytrophanov(1976)]{Dolginov76} Dolginov,
  A.Z., \& Mytrophanov, I.G., 1976, Astrophys. Space Sci., 43, 291

\bibitem[Draine \& Weingartner(1996)]{Draine96} Draine, B.T., \&
  Weingartner, J.C., 1996, \apj, 470, 551

\bibitem[Fornasier et al.(2015))]{Fornasier15} Fornasier, S.,
  Hasselmann, P.H., Barucci, M.A., et al., 2015, \aap, 583, 30

\bibitem[Fulle et al.(2015)]{Fulle15} Fulle, M., Ivanovski, S.L.,
  Bertini, I., et al., 2015, \aap, 583, A14

\bibitem[Fulle et al.(2016a)]{Fulle16a} Fulle, M., Della Corte, V.,
  Rotundi, A., et al., 2016a, \mnras, 462, 132 

\bibitem[Fulle et al.(2016b)]{Fulle16b} Fulle, M., Marzari, F., Della
  Corte, V., et al., 2016b, \apj, 821, 19

\bibitem[Fulle \& Blum(2017)]{FulleBlum17} Fulle, M., \& Blum,
  J., 2017, \mnras, 469, 39

\bibitem[Fulle et al.(2017)]{Fulle17} Fulle, M., Della Corte, V.,
  Rotundi, A., et al., 2017, \mnras, 469, 45

\bibitem[Fulle et al.(2018)]{Fulle18} Fulle, M., Bertini, I., Della Corte, V.,
  et al., 2018, \mnras, 476, 2835

\bibitem[Gerig et al.(2018)]{Gerig18} Gerig, S.-B., Marschall, R.,
  Thomas, N., et al., 2018, Icarus, 311, 1

\bibitem[Gold(1952)]{Gold52} Gold, T., 1952, \mnras, 112, 215

\bibitem[G\"uttler et al.(2018))]{Guettler18} G\"uttler, C., et al.,
  2018, in preparation

\bibitem[Hoang \& Lazarian(2014)]{HoangLazarian14} Hoang, T., \&
  Lazarian, A., 2014, \mnras, 438, 680

\bibitem[Ivanovksi et al.(2017a))]{Ivanovski17a} Ivanovski, S.L., Della
  Corte, V., Rotundi, A., et al., 2017a, \mnras, 469, 774

\bibitem[Ivanovksi et al.(2017b))]{Ivanovski17b} Ivanovski, S.L.,
  Zakharov, V.V., Della  Corte, V., et al., 2017b, Icarus, 282, 333

\bibitem[Jenniskens(1993)]{Jenniskens93} Jenniskens, P., 1993, \aap, 
274, 653

\bibitem[Kimura et al.(2003))]{Kimura03} Kimura, H., Kolokolova, L.,
  \& Mann, I., 2003, \aap, 407, L5 

\bibitem[Kimura et al.(2006))]{Kimura06} Kimura, H., Kolokolova, L.,
  \& Mann, I., 2006, \aap, 449, 1243

\bibitem[Knapmeyer et al.(2018))]{Knapmeyer18} Knapmeyer, M., Fischer,
  H.-H., Knollenberg, J., et al., 2018, Icarus, 310, 165

\bibitem[Kolokolova \& Mackowski(2012))]{KolokolovaMackowski12}
  Kolokolova, L., \& Mackowski, D., 2012,
  J. Quant. Spectrosc. Rad. Transfer, 113, 2567


\bibitem[Kolokolova et al.(2015))]{Kolokolova15} Kolokolova, L., Das,
  H.S., Dubovik, O., et al., 2015, Planet. Space Sci., 116, 30 

\bibitem[Kolokolova et al.(2016))]{Kolokolova16} Kolokolova, L.,
  Koenders, C., Goetz, C., et al., 2016, \mnras, 462, 422

\bibitem[Lasue et al.(2009)]{Lasue09} Lasue, J.,  Levasseur-Regourd,
  A.C., Hadamcik, E., et al. 2009, Icarus, 199, 129

\bibitem[Lazarian(2003))]{Lazarian03} Lazarian, A., 2003,
  J. Quant. Spectrosc. Rad. Transfer, 79, 881

\bibitem[Levasseur-Regourd et al.(1996))]{Levasseur96}
  Levasseur-Regourd, A.C., Hadamcik, E., \& Renard, J.B., 1996, \aap,
  313, 327

\bibitem[Lin et al.(2016)]{Lin16} Lin, Z.-Y., 
Lai, I.L., Su, C.C., et al., 2016, \aap, 588, L3

\bibitem[Macke \& Mishchenko(1996))]{Macke96} Macke, A., \&
  Mishchenko, M.I., 1996, Appl. Optics, 35, 429

\bibitem[Mackowski \& Mishchenko(2011))]{Mackowski11} Mackowski, D.M.,
  \& Mishchenko, M.I., 2011, J. Quant. Spectrosc. Rad. Transfer, 112, 2182

\bibitem[Mannel et al(2017)]{Mannel17} Mannel, T., Bentley, M.S.,
  Boakes, P., et al., 2017, European Planetary Science Congress,
  id. EPSC2017-258


\bibitem[Marschall et al(2016)]{Marschall16} Marschall, R., Su, C.C.,
  Liao, Y., et al., 2016, \aap, 589, A90


\bibitem[Mishchenko et al.(1996))]{Mishchenko96} Mishchenko, M.I.,
  Travis, L.D., \& Mackowski, D.W., 1996,
  J. Quant. Spectrosc. Rad. Transfer, 55, 535 

\bibitem[Mishchenko et al.(2007))]{Mishchenko07} Mishchenko, M.I.,
  Liu, L., Mackowski, D.W., et al., 2007,
  Opt. Express, 15, 2822

\bibitem[Moreno et al.(2007))]{Moreno07} Moreno, F., Mu\~noz, O.,
  Guirado, D., et al., 2007, J. Quant. Spectrosc. Rad. Transfer, 106, 348

\bibitem[Moreno et al.(2017))]{Moreno17} Moreno, F., Mu\~noz, O.,
  Guti\'errez, P.J., et al., 2017, \mnras, 469, 186

\bibitem[Mu\~noz et al.(2017))]{Munoz17} Mu\~noz, O., Moreno, F.,
  Vargas-Mart\'\i n, F., et al., 2017, \apj, 846, 85
  
\bibitem[Rosenbush et al.(2007))]{Rosenbush07} Rosenbush, V.,
  Kolokolova, L., Lazarian, A., et al., 2007, Icarus, 186, 317


\bibitem[Stephens \& Gustafson(1991)]{StephensGustafson91} Stephens,
  J.R., \& Gustafson, B.A.S., 1991, Icarus, 94, 209

\bibitem[Zakharov et al.(2018))]{Zakharov18} Zakharov, V.V., Ivanovski,
  S.L., Crifo, J.F., et al., 2018, Icarus, 312, 121

\bibitem[Zubko et al.(2016))]{Zubko16} Zubko, E., Videen, G., Hines,
  D.C., et al., 2016, Planet. Space. Sci., 123, 63


\end{thebibliography}
\end{document}